\documentclass[%
 reprint,
 amsmath,amssymb,
 aps,
 prd,
floatfix,
]{revtex4-2}

\usepackage{appendix}
\usepackage{placeins}
\usepackage{graphicx}% Include figure files
\usepackage{dcolumn}% Align table columns on decimal point
\usepackage{bm}% bold math
\usepackage{float}
\usepackage{amssymb}
\usepackage{lipsum}
\usepackage{soul}
\usepackage{slashed}
\usepackage{amsmath}
\usepackage{bbold}
\usepackage[compat=1.0.0]{tikz-feynman}
\usetikzlibrary{decorations.markings,decorations.pathreplacing}

\tikzfeynmanset{
	fermion2/.style={
		/tikz/postaction={
			/tikz/decorate=false,
		},
	},
}
\tikzfeynmanset{
	fermion3/.style={
		/tikz/decoration={
			markings,
			mark=at position 0.5 with {
				\node[
				dot,
				fill,
				draw=none,
				] { };
			},
		},
		/tikz/postaction={
			/tikz/decorate=true,
		},
	},
}

\begin{document}

\title{Finite Temperature Isoscalar Instabilities \\ and Moat Regimes in Relativistic Nuclear Matter}

\author{Randall H. V. Pradinett}
\email{randall.vargas@unesp.br}
\author{Theo F. Motta}
\email{theo.motta@unesp.br}
\author{Gastão Krein}
\email{gastao.krein@unesp.br}
\affiliation{Instituto de Física Teórica, Universidade Estadual Paulista, Rua Dr. Bento Teobaldo Ferraz, 271 - Bloco II - 01140-070 São Paulo, SP, Brazil}

\date{\today}

\begin{abstract}
{We study the finite-temperature isoscalar collective response of symmetric nuclear matter in three relativistic mean-field descriptions: the linear Walecka model, its nonlinear extension, and the quark-meson coupling (QMC) model. The Walecka models are formulated in terms of nucleonic degrees of freedom, whereas QMC incorporates quark degrees of freedom confined inside the nucleon. These models differ in their treatment of scalar interactions at finite density, affecting instability boundaries. Focusing on the coupled scalar-vector channel, we track how the boundaries of homogeneous instabilities evolve with temperature. In all three models, we find a nonmonotonic momentum dependence of the static response, characterized by a minimum at finite momentum, a~moat regime that selects a preferred length scale for density fluctuations. Our results show that this moat structure is robust against thermal effects across these relativistic nuclear matter models, providing motivation for future studies of inhomogeneous phases in hot and dense astrophysical matter.}
\end{abstract}

\maketitle

\section{Introduction}
Relativistic nuclear matter has been the focus of intense research for several decades. One of the most resilient gaps in our knowledge has to do with the phase structure of dense matter. The most challenging aspect is determining where the different model-theoretical descriptions remain applicable and where they begin to break down. For instance, at asymptotically large densities, descriptions in terms of baryons interacting through meson exchange is eventually expected to lose its validity completely, as quark and gluon degrees of freedom become relevant. Quantum chromodynamics (QCD) is the most likely theory to govern the fundamental strong force, and thus, perturbative QCD should be adequate to describe dense matter at large densities. For much lower densities, relativistic mean field (RMF) models seem to describe well bulk properties of nuclear matter around saturation, such as density, binding energy, and symmetry energy. Where one effective description fades out of validity is unclear. Essentially, one can say that at low densities the nucleons should behave as point-like particles whereas, as the densities increase, the internal structure of the baryons play greater and greater roles until a point where any effective description ceases to make sense and the only reasonable degrees of freedom are the fundamental quarks and gluons. A key issue regarding the applicability of RMF descriptions for, say, not-so-low densities, has been observed and discussed extensively in the literature \cite{chin1977,Serot:1991st,Furnstahl:1988yj,Perry:1987uu,Lim:1989ma}, namely, the existence of instabilities of the homogeneous ground state of nuclear matter. In Ref.~\cite{Serot:1991st}, a seminal review by Serot and Walecka, they claim that either this instability is an artifact of the approximations, or that ``the composite structure of the baryon must be taken into account''. In other words, are these instabilities a signal that the point-like nucleon description has broken down completely, or is there really the tendency of nuclear matter to form crystalline structures at such high densities?

Recently, the discussion of inhomogeneous phases of dense quark matter \cite{Alford:2000ze,Nickel:2009wj,Buballa:2014tba,Motta:2024rvk,daSilva:2026plc,Buballa:2018hux,Pawlowski:2025jpg} has given a new perspective into this issue. Several works using low-energy models of QCD have shown the energetic preference for inhomogeneous quark matter phases over homogeneous ones \cite{Buballa:2014tba,Buballa:2020nsi,Buballa:2020xaa,Nickel:2009ke,Nickel:2009wj,daSilva:2026plc}. This tendency has subsequently been confirmed via functional methods in QCD \cite{Motta:2024agi,Pawlowski:2025jpg}. Back to baryons, though,
if the homogeneous ground state of \textit{nuclear} matter is unstable against spatially non-trivial density perturbations, its ground state is crystalline \cite{Pitsinigkos:2023xee,Evans:2023hms,Heinz:2013hzanm,Takeda:2018ldi}.
In Ref.~\cite{MottaKrein2025} the authors attempted to answer the question alluded to in the previous paragraph: are these instabilities real, or are they an artifact of point-like nucleons? In \cite{MottaKrein2025} these instabilities were pushed to much higher density in models which take the baryon structure into account. However, non-trivial behavior of correlation functions were much more  consistent amongst the different models, in particular the so-called moat-regime \cite{Rennecke:2025kub,Rennecke:2026ziy,Fu:2024rto}, when static correlation functions are non-monotonic in momentum space.
In this paper, we revisit this discussion at finite temperature, motivated
by the relevance of hot and dense nuclear matter to astrophysical
environments such as core-collapse supernovae and neutron-star mergers.
We investigate how the instabilities found at zero temperature evolve as
the temperature increases and whether the characteristic behavior of the
correlation functions remains robust against thermal effects. To this end,
we consider three different descriptions of nuclear matter:
the linear Walecka model, its nonlinear extension, and the quark--meson coupling (QMC) model, which incorporates the internal structure of the baryon \footnote{The linear and non-linear Walecka models are well introduced and reviewed in Refs.~\cite{Walecka:1995mi,Serot:1991st,Serot:1992ti}. For the QMC model readers should refer to Refs.~\cite{Rikovska-Stone:2006gml,Motta:2022nlj,Guichon:2018uew,guichon1988,guichon1996}.}.

\section{Relativistic Models of Nuclear Matter}\label{rmf}

Within RMF theory, the nuclear interaction is described through the exchange of scalar and vector mesons~\cite{walecka1974,serot1986}. Medium correlations beyond the mean-field approximation are incorporated through the random phase approximation (RPA)~\cite{chin1977}, which provides the polarization functions describing the collective response of the system.

Although all three models we will deal with in this publication describe nuclear interactions similarly, i.e., through scalar and vector mesons, they differ in how medium effects are incorporated.

In the linear Walecka model, nucleons are treated as point-like particles interacting through meson fields with constant couplings. Medium effects come solely through quantum and thermal fluctuations of these fundamental fields, e.g., meson correlation functions are dressed with nucleon-nucleon hole fluctuations. This is a reasonable description, however, although it succeeds reproducing some bulk quantities of nuclear matter, some other quantities come notoriously large, as the nuclear incompressibility.
The nonlinear Walecka model is similar to the above, nucleons are still seen as point-like. However, it extends this description by including self-interactions of the scalar field, which soften the equation of state and modify the density dependence of the effective nucleon mass~\cite{boguta1977}. This correction causes the nucleon mass to become larger at high densities, compared with the linear Walecka model. As a consequence, this model obtains much more reasonable values for the incompressibility, although still somewhat larger than empirical values.
Finally, in the QMC model, meson fields couple directly to the confined quarks inside the nucleon. As a result, the internal structure of the nucleon changes with density, leading to an effective density dependence of the scalar interaction~\cite{guichon1988,guichon1996}. This model aims at taking one modest step towards the description of baryons as confined three quark states. This modest step turns out to be extremely powerful; the model shows impressively good results for the structure of large nuclei \cite{Stone:2016qmi,thomas2022powerfulnewenergydensity}, and succeeds at describing observed neutron star masses with hyperons in the core \cite{Rikovska-Stone:2006gml,Motta:2019tjc,Motta:2022nlj,Leong:2023lmw,Leong:2023yma,Leong:2025fde}. The parameter sets employed in the three models are summarized in
Appendix~\ref{app:model_parameters}.

Throughout this work, we restrict the analysis to the isoscalar sector,
since our aim is to study the finite-temperature evolution of the
instability and moat structure associated with the coupled scalar vector
response of symmetric nuclear matter. Furthermore, the isoscalar and isovector channels do not mix at this level of approximation, and they can be studied separately. In the isoscalar sector, the scalar and
vector channels are coupled through the polarization matrix, whose
eigenvalues determine the collective excitations of the medium. From
their momentum dependence, we determine the instability boundaries and
the emergence of moat regimes at finite temperature.

\subsection{Linear Walecka model}

The linear Walecka model\cite{walecka1974,serot1986}, in its simplest form, describe nucleon interactions as mediated by an isoscalar scalar $\sigma$ meson, responsible for the attractive part of the nuclear force, and an isoscalar vector $\omega^\mu$ meson, which generates the short-range repulsion. The corresponding Lagrangian density is
\begin{align}
\mathcal{L}
&= \bar{\psi}\left[ i\gamma^\mu \partial_\mu - M + g_\sigma \sigma - g_\omega \gamma^\mu \omega_\mu\right]\psi
\nonumber\\
&\quad
+ \frac{1}{2} \left( \partial_\mu \sigma \partial^\mu \sigma - m_\sigma^2 \sigma^2 \right) - \frac{1}{4} F_{\mu\nu}F^{\mu\nu}
\nonumber\\
&\quad + \frac{1}{2} m_\omega^2 \omega_\mu \omega^\mu ,
\label{eq:walecka_lagrangian}
\end{align}
where
\begin{equation}
    F_{\mu\nu} = \partial_\mu \omega_\nu - \partial_\nu \omega_\mu .
\end{equation}
At mean-field level, the scalar field modifies the nucleon mass,
\begin{equation}
M^\ast = M-g_\sigma\bar{\sigma},
\label{eq:effective_mass_lw}
\end{equation}
where the scalar mean field satisfies
\begin{equation}
m_\sigma^2\bar{\sigma}=g_\sigma\rho_s .
\label{eq:sigma_field_lw}
\end{equation}
At finite temperature, the scalar density is
\begin{equation}
\rho_s = \gamma\int\frac{d^3k}{(2\pi)^3} \frac{M^\ast}{E^\ast(k)}
\left[n(k)+\bar n(k)\right],
\label{eq:scalar_density_lw}
\end{equation}
with
\begin{equation}
E^\ast(k)=\sqrt{k^2+M^{\ast 2}},
\end{equation}
and
\begin{equation}
n(k)= \frac{1}{e^{[E^\ast(k)-\mu^\ast]/T}+1},
\qquad
\bar n(k)= \frac{1}{e^{[E^\ast(k)+\mu^\ast]/T}+1},
\label{eq:thermal_distributions_lw}
\end{equation}
are the nucleon and antinucleon occupation
numbers, respectively, and $\gamma=4$ for symmetric nuclear matter.

For fixed $T$ and $\rho_B$, the effective mass $M^\ast$ and the effective
chemical potential $\mu^\ast$ are determined self-consistently from the
scalar-field equation and the baryon-density constraint,
\begin{equation}
\rho_B = \gamma\int\frac{d^3k}{(2\pi)^3} \left[n(k)-\bar n(k)\right].
\label{eq:baryon_density_lw}
\end{equation}
The vector mean field gives
\begin{equation}
m_\omega^2\bar{\omega}_0=g_\omega\rho_B,
\end{equation}
and shifts the chemical potential according to
\begin{equation}
\mu^\ast=\mu-g_\omega\bar{\omega}_0 .
\label{eq:vector_field_lw}
\end{equation}
Temperature dependence of the scalar field, and therefore of
$M^\ast$, enters through the thermal occupation numbers in
Eq.~(\ref{eq:scalar_density_lw}).
The scalar and vector mean fields therefore determine both the in-medium
nucleon mass and the effective chemical potential. Their competition
provides the mechanism for nuclear saturation in the linear Walecka
model. However, the model predicts a relatively large incompressibility
and a low effective nucleon mass at saturation density, resulting in an
equation of state that is generally stiffer than empirical constraints
suggest~\cite{Stone:2014wza}. These limitations motivate the inclusion
of nonlinear meson interactions, which we discuss in the next subsection.

\subsection{Nonlinear Walecka model}

The nonlinear Walecka model extends the linear model by introducing
self-interactions of the scalar meson field~\cite{boguta1977,serot1986}.
These additional terms are contained in the scalar potential
\begin{equation}
U(\sigma) = \frac{b}{3}M(g_\sigma\sigma)^3 +
\frac{c}{4}(g_\sigma\sigma)^4,
\end{equation}
where $b$ and $c$ are phenomenological parameters adjusted to reproduce
the properties of nuclear matter.

The nonlinear terms modify the density dependence of the scalar mean
field and improve the description of nuclear matter around saturation.
In particular, they reduce the large incompressibility of the linear
Walecka model, increase the effective nucleon mass at saturation, and
lead to a softer equation of state~\cite{boguta1977,serot1986}. The
scalar-field equation now becomes
\begin{equation}
m_\sigma^2\bar{\sigma} + bM g_\sigma^3\bar{\sigma}^{\,2} +
c g_\sigma^4\bar{\sigma}^{\,3} = g_\sigma\rho_s ,
\label{eq:sigma_field_nlw}
\end{equation}
while the relation between the scalar field and the effective nucleon
mass remains
\begin{equation}
M^\ast=M-g_\sigma\bar{\sigma}.
\label{eq:effective_mass_nlw}
\end{equation}
Combining Eqs.~(\ref{eq:sigma_field_nlw}) and
(\ref{eq:effective_mass_nlw}) gives
\begin{equation}
M^\ast = M-\frac{g_\sigma^2}{m_\sigma^2}\rho_s + \frac{g_\sigma^2}{m_\sigma^2} \left[ bM(M-M^\ast)^2 + c(M-M^\ast)^3 \right].
\label{eq:effective_mass_nlw_self}
\end{equation}
The scalar density $\rho_s$ is given by
Eq.~(\ref{eq:scalar_density_lw}), with the same finite-temperature
occupation numbers introduced in the previous subsection. The temperature
dependence of the effective mass therefore enters through $\rho_s$, which
determines the scalar mean field self-consistently through
Eq.~(\ref{eq:sigma_field_nlw}). The vector sector remains unchanged from
the linear Walecka model.

\subsection{Quark--meson coupling model}

In the QMC model, nucleons are described as
composite systems of confined quarks rather than point-like particles.
The meson fields couple directly to the quarks inside the nucleon,
allowing its internal structure to respond self-consistently to the
surrounding nuclear medium. As the mean fields change with density, they
modify the quark wave functions and, consequently, the properties of the
nucleon. This provides a microscopic mechanism for the medium dependence
of the nuclear interaction within a relativistic mean-field framework
\cite{guichon1988,saito1997,stone2016}.

Within the MIT bag description, the quark dynamics in the nuclear medium is governed by
\begin{equation}
\left[ i\gamma^\mu \partial_\mu - \left( m_q-g_\sigma^q\bar{\sigma} \right)
- g_\omega^q \gamma^0 \bar{\omega} \right] \psi_q = 0,
\end{equation}
where $g_\sigma^q$ and $g_\omega^q$ denote the quark couplings to the scalar and vector meson fields, respectively. The scalar field modifies the effective quark mass,
\begin{equation}
m_q^\ast = m_q - g_\sigma^q \bar{\sigma},
\end{equation}
while the vector field shifts the quark energies through the vector potential.

The quarks are confined inside a spherical bag of radius $R_N$. Confinement is imposed through the MIT bag boundary condition
\begin{equation}
i\gamma\cdot\hat n\,\psi_q=\psi_q,
\qquad
(r=R_N),
\end{equation}
which leads to a discrete spectrum of allowed quark states. The corresponding eigenenergies are
\begin{equation}
\epsilon_q = \frac{\Omega_q}{R_N} + g_\omega^q \bar{\omega},
\qquad \Omega_q = \sqrt{x_q^2+(m_q^\ast R_N)^2},
\end{equation}
where the eigenvalue $x_q$ is determined by the boundary condition at the bag surface
\cite{guichon1996,saito1997}.

The effective nucleon mass is obtained from the energy of the confined quarks together with the bag volume energy and additional corrections. Schematically,
\begin{equation}
M_N^\ast = E_{\rm quark} + E_{\rm bag} + E_{\rm corr},
\end{equation}
where
\begin{equation}
E_{\rm quark} = \sum_q \frac{\Omega_q}{R_N}, \qquad E_{\rm bag}
=
\frac{4\pi}{3}R_N^3 B_N .
\end{equation}
Here $E_{\rm quark}$ is the energy of the confined quarks, $E_{\rm bag}$ is the vacuum energy associated with the finite bag volume, and $E_{\rm corr}$ contains additional contributions such as center-of-mass and gluon-exchange corrections. These corrections ensure that the free nucleon properties are properly reproduced before the model is extended to nuclear matter
\cite{stone2007,stone2016}.

Since the scalar field modifies the quark wave functions inside the
nucleon, the effective nucleon mass becomes a nonlinear function of
$\bar{\sigma}$. This dependence is well approximated by
\begin{equation}
M^\ast(\bar{\sigma}) = M-g_\sigma\bar{\sigma} +\frac{d_\sigma}{2}(g_\sigma\bar{\sigma})^2,
\end{equation}
where $d_\sigma$ is the scalar polarizability of the nucleon. It
characterizes the response of the nucleon's internal quark structure to
the scalar mean field~\cite{guichon1996,guichon2008}. For the bag
parametrization used here,
\begin{equation}
\frac{d_\sigma}{2} = 0.0022\,\mathrm{fm} +0.1055\,R_N
-0.0178\,\mathrm{fm}^{-1}R_N^2,
\end{equation}
where $R_N$ is expressed in fm.

The corresponding effective scalar coupling is
\begin{equation}
\Gamma_\sigma(\bar{\sigma}) = - \frac{\partial M^\ast}{\partial\bar{\sigma}} = g_\sigma - g_\sigma^2 d_\sigma \bar{\sigma}.
\end{equation}
%---
In the present calculation, temperature does not enter explicitly at the
level of the confined quarks. Instead, finite-temperature effects enter
through the thermal occupation of the nucleon and antinucleon states.
The scalar field is determined self-consistently from
\begin{equation}
m_\sigma^2\bar{\sigma}
=
\Gamma_\sigma(\bar{\sigma})\,\rho_s ,
\label{eq:qmc_sigma_field}
\end{equation}
where $\rho_s$ is the finite-temperature scalar density defined in
Eq.~(\ref{eq:scalar_density_lw}). The thermal occupation numbers therefore
modify $\rho_s$ and, through the equation above, the self-consistent
scalar field.
%---

An important difference from the Walecka models is that, in QMC, the
scalar coupling changes with the scalar field as a consequence of the
internal structure of the nucleon. As the scalar field increases with
density, it modifies the quark wave functions inside the nucleon and,
consequently, its response to the surrounding medium. The resulting
decrease of $\Gamma_\sigma(\bar{\sigma})$ weakens the scalar attraction.
This also gives rise to effective many-body forces: the interaction
between two nucleons depends on the surrounding medium, since the scalar
field produced by the other nucleons changes their internal structure.
In this way, the scalar response of the nucleon provides an effective
many-body repulsion that limits the growth of the scalar attraction at
high density and plays an important role in the saturation of nuclear
matter~\cite{saito1997,guichon2008,stone2016}.

\section{Instabilities and Moat Regimes}
\label{sec:instabilities_moat}

The stability of homogeneous nuclear matter is determined by its response
to small density fluctuations. In the random phase approximation (RPA),
this response is described by the dressed meson propagator,
\begin{equation}
D(q) = \left[ D_0^{-1}(q)-\Pi(q) \right]^{-1},
\end{equation}
where $D_0(q)$ is the free meson propagator and $\Pi(q)$ is the
polarization matrix. The collective modes of the system correspond to
the poles of the dressed propagator, which satisfy
\begin{equation}
\det\!\left[
D_0^{-1}(q)-\Pi(q)
\right]
=
0.
\label{eq:collective_mode_condition}
\end{equation}
The explicit finite-temperature polarization functions are evaluated
following Ref.~\cite{MottaKrein2025}, while the effective masses and
couplings are provided by the nuclear models discussed in the previous
section.

Throughout this work we consider the static limit,
$q_0=0$, where the response is characterized by the static dielectric
function,
\begin{equation}
\epsilon(q) = \det\!\left[ 1-D_0(q)\Pi(q) \right].
\label{eq:sdf_definition}
\end{equation}
For isotropic matter, the response naturally separates into one
longitudinal mode and two degenerate transverse modes,
\begin{equation}
\epsilon(q) = \epsilon_L(q)\epsilon_T^2(q).
\label{eq:sdf_factorization}
\end{equation}

The dielectric function provides a direct criterion for stability. The
boundary of the unstable region is determined by
\begin{equation}
\epsilon(q)=0,
\label{eq:instability_condition}
\end{equation}
while
\begin{equation}
\epsilon_{L}(q)<0
\quad \text{or} \quad
\epsilon_{T}(q)<0
\end{equation}
signals that the homogeneous phase is unstable. When the zero occurs at
$q\simeq0$, 
it characterizes an instability towards another \textit{homogeneous} phase. In this case this is associated with the liquid--gas spinodal
region. A zero at finite momentum instead indicates that the dominant
unstable mode develops at some characteristic momentum. Such
finite-momentum instabilities are essentially the classic Brazovskii
mechanism, which favors the formation of spatially modulated phases
\cite{Brazovskii1975,Nickel:2009ke,Nickel:2009wj}.

Besides locating the instability boundaries, it is also useful to
analyze the momentum dependence of the collective modes. After
diagonalizing the matrix
\begin{equation}
D_0^{-1}(q)-\Pi(q),
\end{equation}
we obtain three independent eigenvalues,
$\lambda_T(q)$, $\lambda_-(q)$, and $\lambda_+(q)$.
A moat regime is identified when one of these eigenvalues develops a
minimum at finite momentum,
\begin{equation}
\frac{\partial \lambda_i(q)}{\partial q}=0,
\qquad
\frac{\partial^2 \lambda_i(q)}{\partial q^2}>0.
\label{eq:moat_condition}
\end{equation}
Unlike the instability criterion, which is determined by the dielectric
function, the moat regime reflects the appearance of a preferred
momentum scale in the collective response. Such finite-momentum minima
have also been discussed in dense QCD and related systems
\cite{RenneckePisarski2021,RenneckePisarski2023}.

As a benchmark, we first verified that the zero-temperature calculation
reproduces the previously reported instability boundaries and moat
structure. We then extended the analysis to finite temperature. As the temperature increases, the region in which a finite-momentum minimum occurs is displaced toward higher baryon densities, while the depth of the minimum decreases. Nevertheless, the moat structure remains visible over a broad temperature range.

\section{Results}

The finite-temperature collective response of the three relativistic models is characterized through the static dielectric function and its eigenvalues. Together, these quantities determine the stability of the homogeneous phase and reveal the emergence of finite-momentum collective modes.

\subsection{Instability regions in the three models}

\begin{figure*}[t]
\centering
\includegraphics[width=0.97\textwidth]{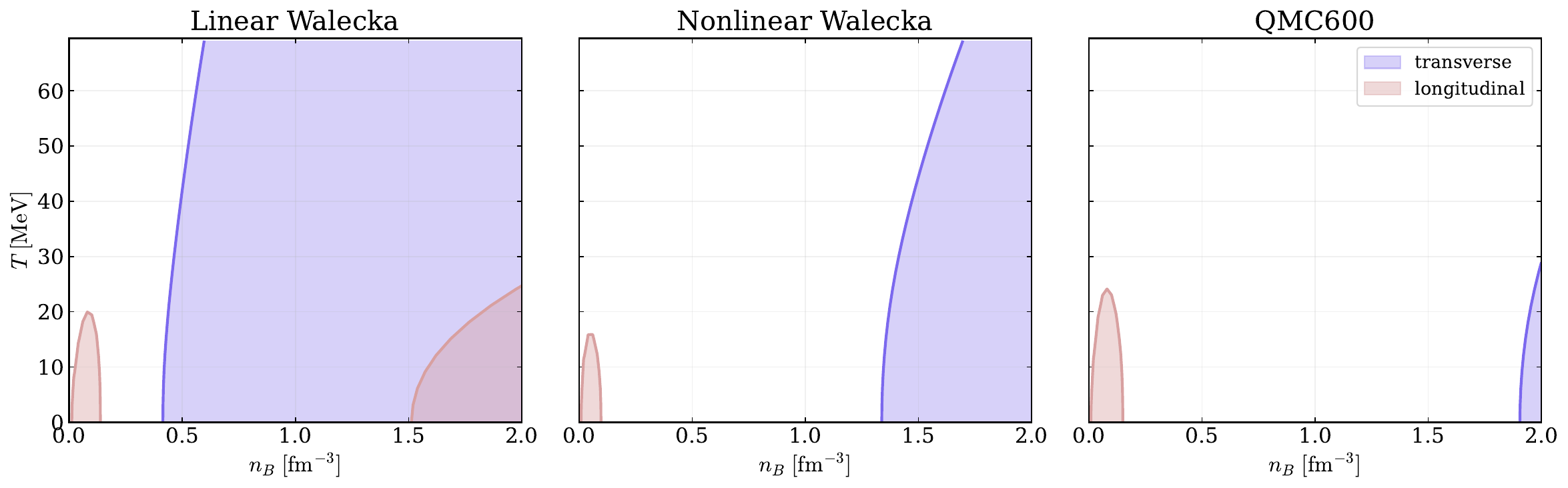}
\caption{
Instability boundaries in the $(T,n_B)$ plane for the linear Walecka, nonlinear Walecka, and QMC models. The shaded regions indicate where the static dielectric function develops zeros.
}
\label{fig:instability-boundaries}
\end{figure*}

Figure~\ref{fig:instability-boundaries} summarizes the instability boundaries in the $(T,n_B)$ plane obtained from the zeros of the static dielectric function for the linear Walecka, nonlinear Walecka, and QMC models. In all three cases, the unstable region spans the whole temperature range, but remains restricted to densities higher than some threshold. For the nonlinear Walecka model and QMC, this threshold is found in a region where the models are not applicable, however, for the linear Walecka model the threshold is substantially smaller. Increasing the temperature shifts the onset of the instability toward higher densities, progressively reducing the density interval over which homogeneous matter becomes unstable.

The low-density longitudinal instability observed in all three models is associated with the liquid--gas phase transition of nuclear matter. In this region, the homogeneous phase separates into phases with different densities, signaling a first-order phase transition. Although the three models display the same qualitative behavior, they predict different critical temperatures. The QMC model yields the highest value of $T_c$, followed by the linear Walecka model, whereas the nonlinear Walecka model predicts a lower critical temperature that lies closer to current empirical estimates. These differences reflect the distinct treatment of medium effects and illustrate how modifications of the effective interaction influence the location of the liquid--gas critical point.

More pronounced differences emerge in the transverse sector. The linear Walecka model exhibits the largest transverse instability region. Including scalar self-interactions in the nonlinear Walecka model modifies the density dependence of the effective nucleon mass and substantially reduces both the location and extent of this region. In the QMC model, the instability domain becomes even more restricted because the scalar polarizability weakens the effective scalar coupling as the density increases. In other words, the medium modifications of the internal nucleon structure play an important role in determining the stability boundaries at intermediate and high densities.

\subsection{Momentum-space moat behavior}

The eigenvalues of the inverse dressed propagator,
\begin{equation}
    D_0^{-1}-\Pi,
\end{equation}
provide additional information beyond the instability boundaries by revealing how the collective response evolves with momentum.

\begin{figure}[t!]
\centering
\includegraphics[width=0.48\textwidth]{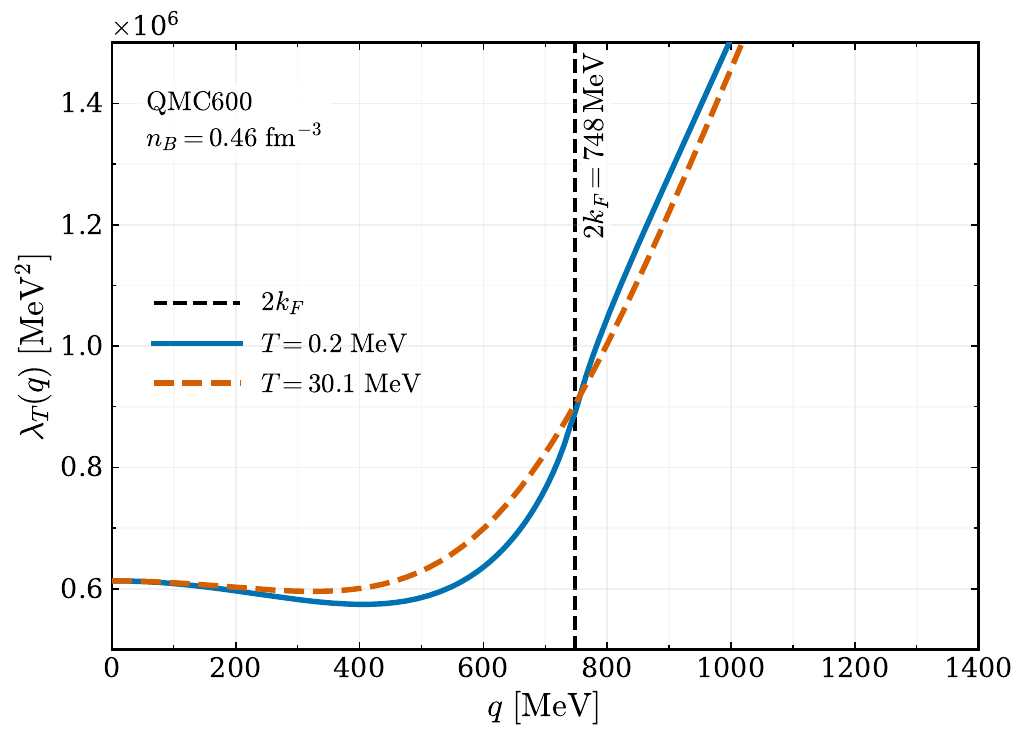}
\caption{
Momentum dependence of the transverse eigenvalue $\lambda_T$ at fixed density for different temperatures. The vertical dashed line indicates the scale $2k_F$ for symmetric nuclear matter at $n_B=0.46~\mathrm{fm}^{-3}$.
}
\label{fig:momentum-moat}
\end{figure}

At low temperature, the eigenvalues develop a non-monotonic momentum dependence with a minimum at finite momentum. Such a minimum is the defining signature of a \textit{moat regime}, indicating that the dominant collective fluctuations occur at a finite wavelength rather than in the long-wavelength limit, as can be seen in Fig.~\ref{fig:momentum-moat}.
The characteristic scale of $2k_F$ marks the wave-number of the Friedel oscillations. . This is a strictly low-temperature phenomenon that manifests itself as an inflection point in the correlation function at $2k_F$, as can also be clearly seen in Fig.~\ref{fig:momentum-moat}. Friedel oscillations are somewhat trivial and simply indicate the sharpness of a Fermi surface. They lead to algebraically decaying oscillations in the correlation functions, as will be discussed shortly. 

As the temperature increases, the finite-momentum minimum becomes less pronounced at fixed density. Rather than disappearing, however, the moat regime is displaced toward higher baryon densities, consistently with the instability boundaries shown in Fig.~\ref{fig:instability-boundaries}.

\subsection{Transverse mode and emergence of a finite-momentum scale}

\begin{figure*}[t]
\centering
\includegraphics[width=\textwidth]{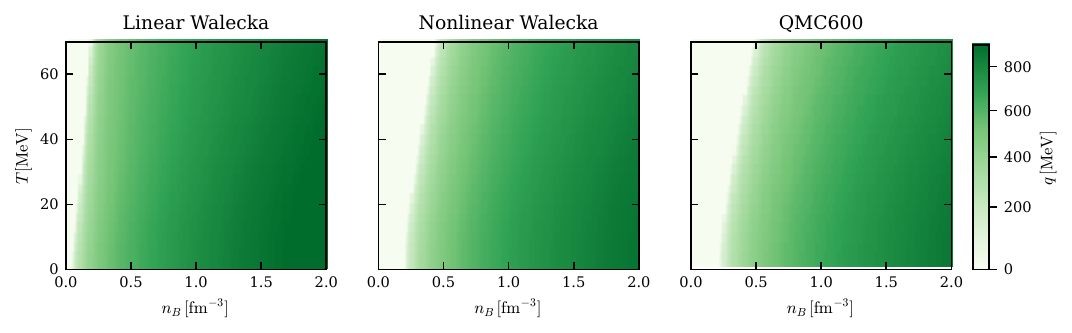}
\caption{
Momentum $q$ at which the transverse mode $\lambda_T(q)$ reaches its minimum, shown as a function of baryon density and temperature for the linear Walecka, nonlinear Walecka, and QMC models.
}
\label{fig:transverse-response}
\end{figure*}

The momentum $q$ at which the transverse eigenvalue $\lambda_T(q)$ reaches its minimum defines the moat  in this channel. Figure~\ref{fig:transverse-response} shows how this momentum evolves throughout the $(T,n_B)$ phase diagram for the three relativistic models. Explicitly,
\begin{equation}
    q(T,n_B) = \underset{q}{\mathrm{arg\,min}} \; \lambda_T(q;T,n_B).
\end{equation}

Since $\lambda_T(q)$ is an eigenvalue of the inverse dressed propagator, $q$ identifies the momentum at which the transverse response is strongest.

The color map illustrates how this characteristic momentum evolves throughout the phase diagram. Small values of $q$ correspond to minima located near zero momentum, whereas larger values indicate that the preferred momentum shifts to increasingly shorter wavelengths.

At low densities, the transverse response is dominated by long-wavelength fluctuations and the minimum occurs near $q=0$. Increasing the density shifts this minimum toward finite momentum, revealing the emergence of a preferred momentum scale associated with the moat regime.

The overall behavior is similar in all three models, although quantitative differences remain. The nonlinear Walecka and QMC models display a more rapid displacement of the minimum toward finite momentum, whereas the linear Walecka model exhibits a smoother evolution over a broader density interval.

Temperature also affects the location of the minimum. As the temperature increases, the region where finite values of $q$ occur moves systematically toward higher baryon densities, indicating that progressively denser matter is required for the transverse channel to develop a pronounced finite-momentum minimum.

\subsection{Configuration-space response and Friedel oscillations}

The configuration-space correlation functions are obtained by Fourier transforming the propagator eigenchannels,
\begin{equation}
    G_i(q)=\frac{1}{\lambda_i(q)}.
\end{equation}
This transformation connects the momentum-space structure of the collective modes to their real-space correlations. Details of the numerical implementation, including the smooth window function and the piecewise-linear integration scheme, are presented in Appendix~\ref{app:friedel}.

Figure~\ref{fig:configuration-space} compares the configuration-space correlation functions of the transverse mode, $\lambda_T$, and the two longitudinal mixed modes, $\lambda_-$ and $\lambda_+$ at $n_B=0.46~\mathrm{fm}^{-3}$ for two representative temperatures.

At $T=0.2~\mathrm{MeV}$, all three channels exhibit algebraically damped oscillations characteristic of Friedel oscillations. Their origin can be traced to the finite-momentum structure of the propagators near $2k_F$, which generates long-range oscillatory correlations after Fourier transformation.

At $T=30.1~\mathrm{MeV}$, thermal effects weaken the momentum-space structure around $2k_F$ and Friedel oscillations are therefore strongly suppressed. The remaining contribution is an oscillatory Yukawa correlation, also known as Yukawa oscillations \cite{Diaz-Alonso:1998eva,DiazAlonso:1989up}. Although these higher temperature oscillations are exponentially dampened by the Yukawa term, they introduce a characteristic momentum in the system capable of affecting observables \cite{RenneckePisarski2021,Nussinov:2024erh}.

The real-space correlations provide a complementary perspective on the momentum-space analysis. At low temperature, the propagator eigenchannels exhibit damped Friedel oscillations, reflecting the finite-momentum structure of the collective response. Friedel oscillation terms vanish exponentially with temperature and the correlation functions approach Yukawa oscillations characteristic of moat regimes.

\begin{figure}[t]
\centering
\includegraphics[width=\columnwidth]{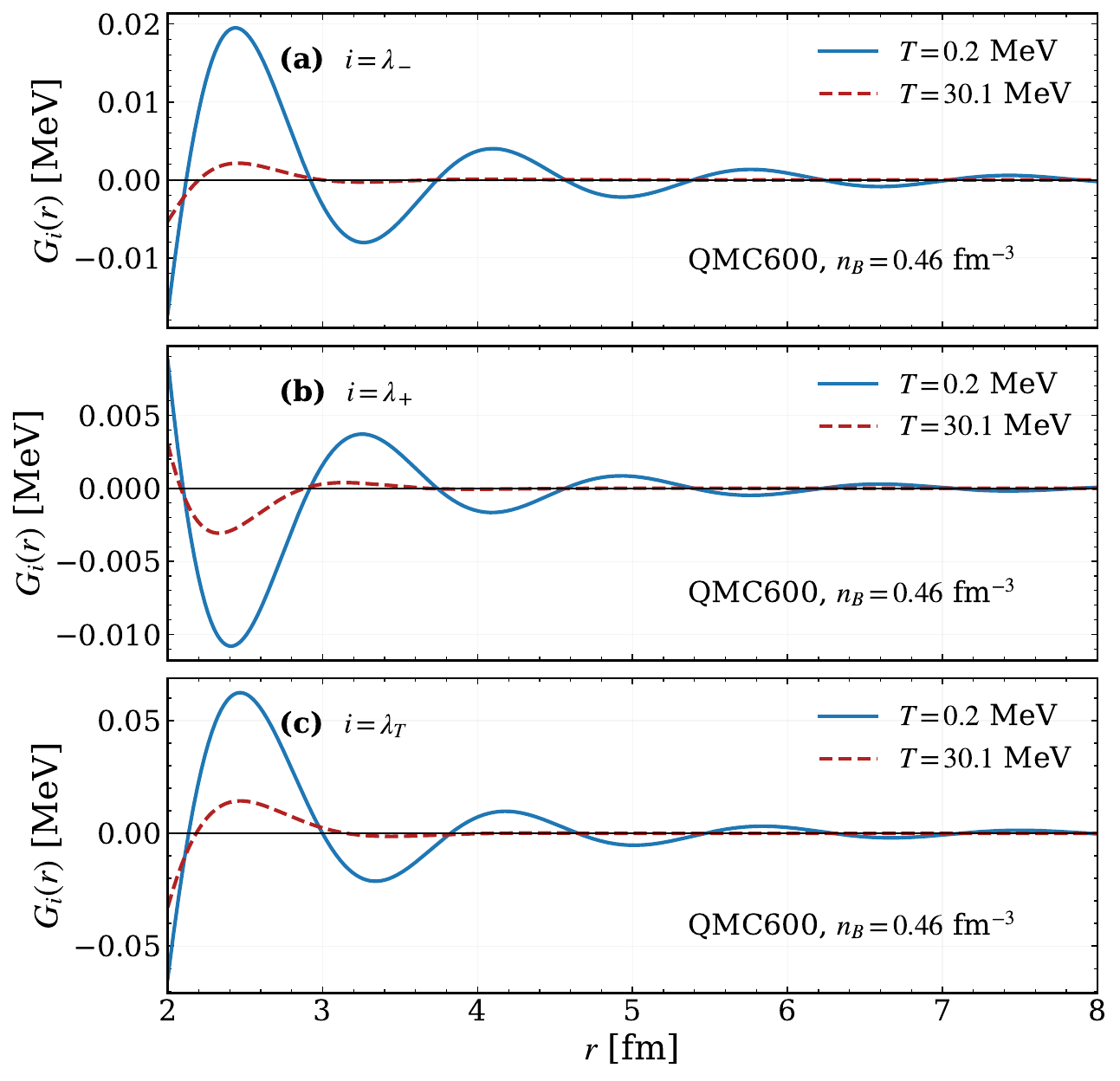}
\caption{
Configuration-space correlation functions for the transverse mode $\lambda_T$ and the longitudinal mixed modes $\lambda_-$ and $\lambda_+$ at $n_B=0.46~\mathrm{fm}^{-3}$ for $T=0.2$ and $30.1~\mathrm{MeV}$.
}
\label{fig:configuration-space}
\end{figure}

\section{Discussion and Conclusions}

{
We have studied the finite-temperature isoscalar collective response of symmetric
nuclear matter in three relativistic mean-field descriptions: the linear Walecka
model, its nonlinear extension, and the QMC model. Focusing on the coupled
scalar--vector sector, we tracked the temperature evolution of the boundaries of
homogeneous instabilities and analyzed the full momentum dependence of the
static response. The main difference among these models lies in their treatment
of scalar interactions at finite density, which directly impacts both the
instability diagram and the emergence of finite-momentum structure.

The linear Walecka model produces the largest unstable regions, including a
high-density longitudinal instability. This feature is strongly reduced in the
nonlinear Walecka model and is further reduced in QMC. One should keep in mind,
however, that the most reliable density range of such hadronic descriptions is
likely limited to $n_B\lesssim 0.5\,\mathrm{fm}^{-3}$ (a few times saturation),
where nucleons with radii of order $0.8\,\mathrm{fm}$ start to overlap.
Nevertheless, the comparison clearly shows that scalar self-interactions and the
internal response of the nucleon soften the collective dynamics and restrict the
range of unstable matter. In this sense, the instability boundaries are highly
sensitive to medium effects.

The momentum dependence of the propagator eigenvalues contains information that
is not captured by instability boundaries alone. In all three models, the
relevant eigenchannels develop a nonmonotonic $q$-dependence with a minimum at
finite momentum. This identifies a moat regime and introduces a preferred length
scale for density fluctuations.

Temperature changes this picture in a systematic way. The finite-momentum
minimum remains present, but its onset is pushed to higher baryon densities and
the minimum becomes less pronounced. Thus, thermal effects do not eliminate the
moat regime, but they require denser matter for the same finite-momentum
structure to develop.

The configuration-space results provide a complementary view of the spatial
correlations. At low temperature, the eigenchannels exhibit damped Friedel
oscillations tied to structure near the Fermi surface. As the temperature increases,
these oscillations are suppressed and the correlations cross over to an
oscillatory Yukawa-like form.

Overall, moat regimes emerge as a robust feature of these relativistic models of
dense nuclear matter, although their location and strength depend on both the
microscopic model and temperature. They may be viewed as an indicator that
inhomogeneous phases could become relevant once finite-momentum correlations are
sufficiently enhanced, even though determining the structure of such phases is
beyond the scope of the present work.

In summary, finite temperature shifts the moat regime toward higher densities
and weakens its momentum dependence while preserving it over a broad region of
the phase diagram. These results provide a basis for future studies of
inhomogeneous phases and their possible role in hot and dense astrophysical
matter.
}

\begin{acknowledgments}
This work was partially financed by Fundação de Amparo à Pesquisa do
Estado de São Paulo (FAPESP), grant Nos. 2024/13426-0 (T.F.M.) and
2018/25225-9 (G.K.); Conselho Nacional de Desenvolvimento Científico e
Tecnológico (CNPq), grant Nos. 315225/2025-4 (T.F.M.) and
309262/2019-4 (G.K.); and Coordenação de Aperfeiçoamento de Pessoal de
Nível Superior (CAPES), grant No. 88887.893649/2023-00 (R.H.V.P.).
\end{acknowledgments}

\appendix

\section{Scalar polarization at finite temperature}
\label{app:scalar_polarization}

The medium-dependent scalar polarization reads
\begin{equation}
\begin{aligned}
\Pi_s(q)={}&-i\gamma_I(\Gamma_s^{\rm eff})^2
\int\frac{d^4k}{(2\pi)^4}\\[-2pt]
&\times{\rm Tr}_D[G(k-l)G(k+l)],\quad q=2l.
\end{aligned}
\end{equation}
Here $\gamma_I=2$ for symmetric matter, $E_k^*=\sqrt{k^2+M^{*2}}$, $\mu^*=\mu-g_\omega\bar\omega_0$, and $n_F(E)=[e^{\beta(E-\mu^*)}+1]^{-1}$, $\bar n_F(E)=[e^{\beta(E+\mu^*)}+1]^{-1}$. The propagator is
\begin{equation}
\begin{aligned}
G(k)&=G_{\rm F}(k)+G_{\rm D}(k),\\
G_{\rm F}(k)&=\frac{\slashed{k}+M^*}{k^2-M^{*2}+i\epsilon},\\
G_{\rm D}(k)&=2\pi i(\slashed{k}+M^*)\delta(k^2-M^{*2})\eta(k_0),\\
\eta(k_0)&=\theta(k_0)n_F(E_k^*)+\theta(-k_0)\bar n_F(E_k^*).
\end{aligned}
\end{equation}
In the no-sea approximation, the vacuum--vacuum term is omitted:
\begin{equation}
\Pi_s=\Pi_s^{\rm FD}+\Pi_s^{\rm DF}+\Pi_s^{\rm DD}.
\label{eq:app_decomposition}
\end{equation}
The common Dirac trace is
\begin{equation}
N(k,l)=4(k^2-l^2+M^{*2}).
\end{equation}

\subsection{FD contribution}
The density insertion on the $k-l$ line gives
\begin{equation}
\begin{aligned}
\Pi_s^{\rm FD}(q)={}&\gamma_I(\Gamma_s^{\rm eff})^2
\int\frac{d^4k}{(2\pi)^4}\\
&\times\frac{2\pi N(k,l)\delta[(k-l)^2-M^{*2}]}{(k+l)^2-M^{*2}+i\epsilon}
\eta(k_0-l_0).
\end{aligned}
\end{equation}
With $E_-^*=\sqrt{(\boldsymbol{k}-\boldsymbol{l})^2+M^{*2}}$,
\begin{equation}
\delta[(k-l)^2-M^{*2}]=\frac{1}{2E_-^*}
\sum_{s=\pm1}\delta(k_0-l_0-sE_-^*).
\end{equation}
Performing the energy integration,
\begin{equation}
\begin{aligned}
\Pi_s^{\rm FD}(q)={}&\gamma_I(\Gamma_s^{\rm eff})^2
\int\frac{d^3k}{(2\pi)^3}\sum_{s=\pm1}
\frac{\eta(sE_-^*)}{2E_-^*}\\
&\times\left.\frac{N(k,l)}{(k+l)^2-M^{*2}+i\epsilon}
\right|_{k_0=l_0+sE_-^*}.
\end{aligned}
\end{equation}
After $\boldsymbol{k}-\boldsymbol{l}\to\boldsymbol{k}$, set $q_0=l_0=0$, $q=|\boldsymbol q|$, and $x=\cos\theta$. Then
\begin{equation}
\begin{aligned}
N(k,l)&=8M^{*2}-4kqx,\\
(k+l)^2-M^{*2}&=-q^2-2kqx,\\
\frac{N(k,l)}{(k+l)^2-M^{*2}}&=2-\frac{2(4M^{*2}+q^2)}{q^2+2kqx}.
\end{aligned}
\end{equation}
The real angular integral is
\begin{equation}
\operatorname{Re}\int_{-1}^1\frac{dx}{q^2+2kqx+i0^+}
=\frac{1}{2kq}\ln\left|\frac{q+2k}{q-2k}\right|.
\end{equation}
The real part follows from the principal-value prescription:
\begin{equation}
\begin{aligned}
\operatorname{Re}\Pi_s^{\rm FD}(0,q)
={}&\frac{\gamma_I(\Gamma_s^{\rm eff})^2}{4\pi^2}
\int_0^\infty\!dk\,\frac{k^2}{E_k^*}\\
&\times[n_F(E_k^*)+\bar n_F(E_k^*)]\\
&\times\left[2-\frac{4M^{*2}+q^2}{2kq}
\ln\left|\frac{q+2k}{q-2k}\right|\right].
\end{aligned}
\label{eq:app_fd_real}
\end{equation}

\subsection{DF contribution}
With the density insertion on the $k+l$ line,
\begin{equation}
\begin{aligned}
\Pi_s^{\rm DF}(q)={}&\gamma_I(\Gamma_s^{\rm eff})^2
\int\frac{d^4k}{(2\pi)^4}\\
&\times\frac{2\pi N(k,l)\delta[(k+l)^2-M^{*2}]}{(k-l)^2-M^{*2}+i\epsilon}
\eta(k_0+l_0).
\end{aligned}
\end{equation}
Writing $E_+^*=\sqrt{(\boldsymbol{k}+\boldsymbol{l})^2+M^{*2}}$,
\begin{equation}
\delta[(k+l)^2-M^{*2}]=\frac{1}{2E_+^*}
\sum_{s=\pm1}\delta(k_0+l_0-sE_+^*),
\end{equation}
one obtains
\begin{equation}
\begin{aligned}
\Pi_s^{\rm DF}(q)={}&\gamma_I(\Gamma_s^{\rm eff})^2
\int\frac{d^3k}{(2\pi)^3}\sum_{s=\pm1}
\frac{\eta(sE_+^*)}{2E_+^*}\\
&\times\left.\frac{N(k,l)}{(k-l)^2-M^{*2}+i\epsilon}
\right|_{k_0=-l_0+sE_+^*}.
\end{aligned}
\end{equation}
After $\boldsymbol{k}+\boldsymbol{l}\to\boldsymbol{k}$, the static expressions are
\begin{equation}
\begin{aligned}
N(k,l)&=8M^{*2}+4kqx,\\
(k-l)^2-M^{*2}&=-q^2+2kqx,\\
\frac{N(k,l)}{(k-l)^2-M^{*2}}&=2-\frac{2(4M^{*2}+q^2)}{q^2-2kqx}.
\end{aligned}
\end{equation}
Changing $x\to-x$ gives
\begin{equation}
\operatorname{Re}\Pi_s^{\rm DF}(0,q)
=\operatorname{Re}\Pi_s^{\rm FD}(0,q).
\label{eq:app_df_real}
\end{equation}
The logarithmic singularity at $k=q/2$ is integrable.

\subsection{DD contribution}
The two-density term is
\begin{equation}
\begin{aligned}
\Pi_s^{\rm DD}(q)={}&i\gamma_I(\Gamma_s^{\rm eff})^2(2\pi)^2
\int\frac{d^4k}{(2\pi)^4}N(k,l)\\
&\times\delta[(k-l)^2-M^{*2}]
\delta[(k+l)^2-M^{*2}]\\
&\times\eta(k_0-l_0)\eta(k_0+l_0).
\end{aligned}
\end{equation}
The $k_0$ integration leaves
\begin{equation}
\delta(2l_0+sE_-^*-s'E_+^*),\qquad s,s'=\pm1.
\end{equation}
For $l_0=0$ and $q>0$, only $s=s'$ contributes, and $E_-^*=E_+^*$ requires $\boldsymbol{k}\cdot\boldsymbol q=0$. Hence
\begin{equation}
N(k,l)=8(M^{*2}-l^2)=2(4M^{*2}+q^2).
\end{equation}
The angular delta function imposes $k\ge q/2$ and contributes the Jacobian $E^*/(kq)$. With $E^*=\sqrt{k^2+M^{*2}}$,
\begin{equation}
\begin{aligned}
\Pi_s^{\rm DD}(0,q)={}&i\frac{\gamma_I(\Gamma_s^{\rm eff})^2}{4\pi q}
(4M^{*2}+q^2)\\
&\times\int_{E_{\rm min}^*}^\infty\!dE^*\,
[n_F^2(E^*)+\bar n_F^2(E^*)],
\end{aligned}
\label{eq:app_dd_static}
\end{equation}
where $E_{\rm min}^*=\sqrt{M^{*2}+q^2/4}$.

\subsection{Combined scalar polarization}
Combining the real one-density terms,
\begin{equation}
\begin{aligned}
\operatorname{Re}\Pi_s(0,q)={}&
\frac{\gamma_I(\Gamma_s^{\rm eff})^2}{2\pi^2}
\int_0^\infty\!dk\,\frac{k^2}{E_k^*}\\
&\times[n_F(E_k^*)+\bar n_F(E_k^*)]\\
&\times\left[2-\frac{4M^{*2}+q^2}{2kq}
\ln\left|\frac{q+2k}{q-2k}\right|\right].
\end{aligned}
\label{eq:app_scalar_real}
\end{equation}
The on-shell parts of FD and DF also contribute to the time-ordered imaginary part:
\begin{equation}
\begin{aligned}
\operatorname{Im}[\Pi_s^{\rm FD}+\Pi_s^{\rm DF}]
={}&-\frac{\gamma_I(\Gamma_s^{\rm eff})^2}{4\pi q}
(4M^{*2}+q^2)\\
&\times\int_{E_{\rm min}^*}^\infty\!dE^*\,
[n_F(E^*)+\bar n_F(E^*)].
\end{aligned}
\end{equation}
Together with the DD term,
\begin{equation}
\begin{aligned}
\operatorname{Im}\Pi_s^{\rm F}(0,q)
={}&-\frac{\gamma_I(\Gamma_s^{\rm eff})^2}{4\pi q}
(4M^{*2}+q^2)\\
&\times\int_{E_{\rm min}^*}^\infty\!dE^*\,
\bigl\{n_F(E^*)[1-n_F(E^*)]\\
&\qquad+\bar n_F(E^*)[1-\bar n_F(E^*)]\bigr\}.
\end{aligned}
\label{eq:app_scalar_imag}
\end{equation}
Thus
\begin{equation}
\Pi_s^{\rm F}(0,q)=\operatorname{Re}\Pi_s(0,q)
+i\operatorname{Im}\Pi_s^{\rm F}(0,q).
\label{eq:app_scalar_final}
\end{equation}
For fixed $q>0$, the retarded polarization has no zero-frequency imaginary part; the finite-temperature imaginary term above belongs to the time-ordered correlator.

\section{Configuration-space correlations}
\label{app:friedel}

We examine the spatial behavior of the static collective modes through the eigenvalues $\lambda_i(q)$ of the inverse dressed propagator,
\begin{equation}
D^{-1}(0,q)=D_0^{-1}(0,q)-\operatorname{Re}\Pi(0,q),
\end{equation}
where $i=T,-,+$. The corresponding propagator eigenvalues are
\begin{equation}
G_i(q)=\frac{1}{\lambda_i(q)}.
\label{eq:Giq}
\end{equation}
Their Fourier transforms characterize the spatial behavior of these eigenchannels. Since the diagonalizing eigenvectors may depend on $q$, the transforms are not, in general, correlators of fixed local meson fields.

For isotropic matter, using the natural-unit momentum-space normalization and expressing $r$ in fm, the radial transform is
\begin{equation}
\begin{aligned}
G_i(r)={}&\frac{\hbar c}{2\pi^2r}\int_0^{q_{\max}}dq\,q\,
\sin\!\left(\frac{qr}{\hbar c}\right) G_i(q)W(q),
\end{aligned}
\label{eq:Gr}
\end{equation}
with $\hbar c=197.3269804\ \mathrm{MeV\,fm}$. At $n_B=0.46\ \mathrm{fm}^{-3}$, the Fermi momentum is
\begin{equation}
k_F=\left(\frac{3\pi^2n_B}{2}\right)^{1/3}
\simeq1.895\ \mathrm{fm}^{-1},
\qquad 2\hbar c k_F\simeq748\ \mathrm{MeV}.
\end{equation}

The input eigenvalues are sampled at 800 momenta over $0.5\leq q\leq4000$ MeV. To suppress endpoint artifacts we use
\begin{equation}
W(q)=\exp\!\left[-\left(\frac{q}{q_c}\right)^p\right],
\quad q_c=1800\ \mathrm{MeV},\quad p=4.
\label{eq:gaussian_window}
\end{equation}
The integrand $qG_i(q)W(q)$ is interpolated linearly between neighboring momentum points, and its product with the sine is integrated analytically over each interval. Independent numerical quadratures reproduce the transforms to better than approximately $0.2\%$ in relative root-mean-square difference. Tests with different momentum cutoffs and window parameters show that the oscillatory profiles are reproducible, although their amplitudes retain a non-negligible dependence on the window.

At low temperature, the oscillations are consistent over the sampled distance range with a Friedel-like envelope,
\begin{equation}
G_i(r)\sim\frac{A_i}{r^{\alpha_i}}
\cos\!\left(2k_Fr+\delta_i\right),
\label{eq:friedel_fit}
\end{equation}
where $k_F$ is expressed in $\mathrm{fm}^{-1}$ and the cosine argument is equivalently $2k_Fr$. At $T=30.1$ MeV the oscillations are substantially suppressed relative to $T=0.2$ MeV. These observations characterize the finite interval studied; neither an exact asymptotic $r^{-3}$ law nor a only Yukawa damping scale is inferred from the fits.

\section{Model parameters}
\label{app:model_parameters}

For completeness, we summarize the parameter sets employed in the
linear Walecka, nonlinear Walecka, and quark--meson coupling (QMC)
calculations. These are the parameters used in the numerical evaluation
of the mean fields, polarization functions, and collective response
discussed in Sec.~IV.

Table~\ref{tab:common_parameters} lists the vacuum nucleon and meson
masses together with the scalar and vector meson--nucleon couplings.
The three models use the same nucleon and vector-meson masses, whereas
the scalar-meson mass and the interaction strengths depend on the
chosen parametrization.

\begin{table}[H]
\squeezetable
\centering
\caption{Masses and meson--nucleon couplings used in the linear
Walecka, nonlinear Walecka, and QMC models.}
\label{tab:common_parameters}
\begin{ruledtabular}
\begin{tabular}{lccc}
Parameter
& Linear
& Nonlinear
& QMC
\\
\hline
$M$ [MeV]
& $939$
& $939$
& $939$
\\
$m_{\sigma}$ [MeV]
& $550$
& $550$
& $600$
\\
$m_{\omega}$ [MeV]
& $783$
& $783$
& $783$
\\
$g_{\sigma}$
& $9.573$
& $8.685$
& $9.02$
\\
$g_{\omega}$
& $11.670$
& $8.646$
& $8.42$
\end{tabular}
\end{ruledtabular}
\end{table}

The additional model-dependent parameters are listed in Table~\ref{tab:additional_parameters}. In the nonlinear Walecka model, the coefficients $b$ and $c$ control the cubic and quartic self-interactions of the scalar field and therefore modify the density
dependence of the effective nucleon mass. In the QMC model, $d_{\sigma}$ is the scalar polarizability of the nucleon, while $R_N$ is the free nucleon bag radius. These quantities describe the response of the internal quark structure of the nucleon to the surrounding scalar field.

\begin{table}[H]
\centering
\caption{Additional parameters entering the nonlinear Walecka and QMC
models. A dash indicates that the corresponding parameter is absent
from that model.}
\label{tab:additional_parameters}
\begin{ruledtabular}
\begin{tabular}{lcc}
Parameter
& Nonlinear Walecka
& QMC
\\
\hline
$b$
& $7.950\times10^{-3}$
& ---
\\
$c$
& $6.952\times10^{-4}$
& ---
\\
$d_{\sigma}$ [fm]
& ---
& $0.15$
\\
$R_N$ [fm]
& ---
& $0.80$
\end{tabular}
\end{ruledtabular}
\end{table}

\bibliographystyle{apsrev4-2}
\bibliography{references}

\end{document}